\documentclass{article}
\def\Tr{{\hbox{Tr}}}
 \def\hx{\hat{x}}
\def\bphi{\bar{\phi}} \def\hphi{\hat{\phi}}
\def\hbphi{\hat{\bar{\phi}}} \def\hsigma{\hat{\sigma}}
\def\2{{\cal O}(2)}
\def\da{{\dot{a}}}
\def\db{{\dot{b}}}
\def\cP{{\cal P}}
\def\cQ{{\cal Q}}
\def\hvac{{\hbox{vac}}}
\def\cW{{\cal W}}
\def\bp{{\bf p}}
\def\bk{{{\bf k}}}
\def\tk{{\tilde{k}}}
\def\bx{{{\bf x}}}

\def\tbk{{\tilde{\bk}}}

\def\tpartial{{\tilde{\partial}}}
\def\bpartial{{\bf \partial}}
\def\tbpartial{{\tilde{\bpartial}}}
\def\fitheta{{\frac{i\theta}2}}
\def\ftheta{{\frac{\theta}2}}
\def\rmi{{\rm i}}
\begin{document}
\begin{flushright}    
UFIFT-HEP-02-23 
\end{flushright}
\vskip 1cm

\centerline{\Large Quantization of Noncommutative Scalar Solitons at finite $\theta$}
\vskip 1cm
\centerline{Xiaozhen Xiong\footnote{email: xiaozhen@phys.ufl.edu}}
\date{}
\vskip 0.5cm
\centerline{\it Institute for Fundamental Theory,}
\centerline{\it Department of Physics, University of Florida}
\centerline{\it Gainesville FL 32611, USA}
\vskip 1cm
\begin{abstract}
We start by discussing the classical noncommutative (NC) $Q$-ball solutions 
near the commutative limit, then generalize the virial relation. 
Next we quantize the NC $Q$-ball canonically. At very small $\theta$ 
quantum correction to the energy of the $Q$-balls is calculated through 
summation of the phase shift. UV/IR mixing terms are found in the 
quantum corrections which cannot be renormalized away. The same method 
is generalized to the NC GMS soliton for the smooth enough solution. 
UV/IR mixing is also found in the energy correction and UV divergence 
is shown to be absent. In this paper only ($2+1$) dimensional scalar field 
theory is discussed. 
\end{abstract}
\vskip 1cm
\noindent PACS number(s): 11.10.-z, 11.10.Gh, 11.30.-j 12.60.Jv
\vskip 1cm
\vskip 1cm

\section{Introduction}

Noncommutative (NC) field theory has been considered as certain limit of 
the effective action of open string modes on the brane~\cite{SW,Connes}. 
Since the discovery of NC solitons in NC scalar field theory 
by Gopakumar, Minwalla and Strominger (GMS)~\cite{GMS}, 
the soliton solutions have been explicitly
constructed in different gauge theories with or without matter~\cite{Nek,Komba}.  
Such NC solitons can be interpreted as lower dimensional $D$-branes in
string field theory~\cite{MinUnstable,HarveyDbrane}. 

GMS solitons, which exist in ($2+1$)
dimensional NC scalar field theory, while classically stable, 
cease to exist at sufficiently small NC parameter $\theta$\ , due to the
nonexistence theorem of Derrick~\cite{Derrick} in the commutative
limit($\theta\rightarrow 0$).  In this commutative limit, however,
time dependent nontopological solitons, or $Q$-balls exist in all
space dimensions~\cite{LeeReport,Coleman}. 
In this paper we'll study the NC generalization of time dependent
$Q$-balls, NC $Q$-balls first. In particular, we found all the stable 
soliton solution
family, which depend smoothly on the NC parameter $\theta$ and have a
closed form in the NC limit ($\theta\rightarrow\infty$)~\cite{Dur},
sustain for arbitrary small $\theta$\ . Classically they
all reduce to the ordinary $Q$-ball solution in the commuative
limit. Thus it becomes interesting to investigate the  
quantum properties of NC solitons near the commutative limit. In the
perturbative NC field theory UV/IR mixing occurs in renormalization~\cite{MRS}. 
We discuss the similar NC effects in the quantization of the
nonperturbative NC field theory.  The same method is further generalized 
to the GMS soliton case at finite $\theta$\ . In this paper we deal, for
simplicity, only with $(2+1)$ dimensional NC scalar field theory.

\section{Classical Noncommutative $Q$-ball Solution}

\subsection{Hamiltonian and Equation of motion}

In this section we derive the equation of motion for NC $Q$-ball
solutions, following brief introduction of NC scalar field
theory. The form of the solution is already given in~\cite{NCQball}. 
We discuss the existence and stablity of the solutions,
and show that in the commutative limit NC $Q$-balls just reduce to the
commutative $Q$-balls. 

Consider a NC scalar field theory action with global $U(1)$ phase
invariance,
\begin{equation}
S=-\int
dtd^2x\left[\partial_\mu\bar{\phi}\partial^\mu\phi+V(\frac 12\{\bar{\phi},\phi\})\right]\
,                                                                                               \label{action}
\end{equation}
where the space-time metric is ($-,+,+$), and the fields are
multiplied by NC star product, generally made implicit in this
paper, and $\{A\ ,B\}\equiv A\star B+B\star A$\ . 
The potential $V$ has a global minimum at the origin, with the scaling property
\begin{equation}
  V(\bar{\phi}\phi)=g^{-2}V(g^2\bar{\phi}\phi)\ .                                         \label{ScalePot} 
\end{equation}
$g$ is then the coupling constant assumed to be small. 
The commutative limit of this action is where ordinary $Q$-balls have
already been constructed~\cite{LeeReport,Coleman}. The NC star product is defined to
be,
\begin{equation}
(\phi\star\psi)(x)\equiv\left.\hbox{exp}(i\frac\theta
2\epsilon_{jk}\frac\partial{\partial x_j}\frac\partial{\partial
y_k})\phi(x)\psi(y)\right|_{y=x}\ ,                                                        \label{Star}  
\end{equation} where $j,k=1,2$\ .
The NC algebra of the functions with the star product defined above is
well known to be isomorphic to an algebra of operators on a one particle Hilbert space~\cite{Zachos},
\begin{equation}
[\hat{x}^1\ ,\ \hat{x}^2]=i\theta\ .
\end{equation}
In this isomorphism, also known as Weyl transform, the star product 
is just the operator product,
\begin{equation}
\phi(x)\star\psi(x)\longleftrightarrow
\hat{\phi}(\hat{x})\hat{\psi}(\hat{x})\ ,
\end{equation}
and
\begin{equation}
\int\phi(x)d^2x=2\pi\theta\mbox{Tr}\hat{\phi}(\hat{x})\ ,
\end{equation}
where $\hat{\phi}(\hat{x})\leftrightarrow\phi(x)$ are 
the Weyl transforms of each other.
The derivative are written as
\begin{equation}
\frac\partial{\partial x^i}\phi(x)\longleftrightarrow \frac
i\theta\epsilon_{ij}[\hat{x}^j,\hat{\phi}(\hat{x})]\ .
\end{equation}
Define creation and annihilation operators
\begin{equation} 
a=\frac 1{\sqrt{2\theta}}(\hat{x}^1+i\hat{x}^2)\ ,\qquad\ a^\dag=\frac
1{\sqrt{2\theta}}(\hat{x}^1-i\hx^2)\ ,
\end{equation}
with $[a,a^\dag]=1$ as usual. The field $\phi(x)$ or $\hphi(\hx)$ 
can be expanded in the orthonormal basis $f_{nm}(x)$ or $|n><m|$~\cite{Komba}. 

In the operator formalism the action
integral (\ref{action}) becomes
\begin{equation}
S[\hat{\phi}, \hat{\bar{\phi}}]=\int
dt2\pi\theta\mbox{Tr}\left(\partial_0\hbphi\partial_0\hphi+\frac
1\theta([a,\hbphi][a^\dag,\hphi]+
[a^\dag,\hbphi][a,\hphi])-V(\hat{\bar{\phi}}\hat{\phi})\right).
\end{equation}
The equation of motion is
\begin{equation}
\partial_0^2\hphi+\frac 2\theta[a,[a^\dag,\hphi]]+\hphi
V(\hbphi\hphi)=0\ .                                                                                 \label{EqnM}
\end{equation}

The action has global $U(1)$ phase invariance, which yields a conserved charge
\begin{equation}
Q[\hphi,\hbphi]=\int d^2x
j^0=i2\pi\theta\mbox{Tr}(\hat{\bar{\phi}}\partial_0
\hat{\phi}-\partial_0\hat{\bar{\phi}}\hat{\phi})\ .                                                  \label{ConCharge}
\end{equation}                                                                                        
$Q$ is interpreted as particle number in the physical system. A
particular system always exists with fixed particle number
$N=Q[\hphi,\hbphi]$\ .  To find nondissipative soliton
solutions~\cite{LeeReport,Coleman} under this constraint, we write Hamiltonian
\begin{equation}
H=2\pi\theta\mbox{Tr}\left(\partial_0\hbphi\partial_0\hphi-\frac
1\theta([a,\hbphi][a^\dag,\hphi]+
[a^\dag,\hbphi][a,\hphi])+V(\hat{\bar{\phi}}\hat{\phi})\right)+\omega(N-Q[\hphi,\hbphi])\ ,
\end{equation}
with the constraint applied before the Poisson bracket is worked out \cite{Dirac}. 
The minimum energy solution occurs at
\begin{equation}
\left.\frac{\delta
H}{\delta(\partial_0\hbphi)}\right|_N=\partial_0\hphi+i\omega\hphi=0\
,
\end{equation}
which yields 
\begin{equation}
\hphi=\frac 1{\sqrt{2}}\hsigma(\hat{x})e^{-i\omega t}.
\end{equation}
Assuming hermitian $\hsigma(\hat{x})$ or real $\sigma(x)$, $H$ becomes
\begin{equation}
H=2\pi\theta\Tr\left(-\frac 12\omega^2\hsigma^2-\frac
1\theta[a,\hsigma][a^\dag,\hsigma]+V(\frac 12\hsigma^2)\right)+\omega
N\ ,\label{NHam}
\end{equation}
with the particle number
\begin{equation}
N=2\pi\theta\omega\Tr(\hsigma^2)\ .\label{N}
\end{equation}
and the equation of motion (\ref{EqnM})
\begin{equation}
\frac 2\theta[a,[a^\dag,\hsigma]]-\omega^2\hsigma+\hsigma V'(\frac
12\hsigma^2)=0\ .\label{NEqnM}
\end{equation}
Note the equation of motion (\ref{NEqnM}) also follows from
$({\delta H}/{\delta \hsigma})|_N=0$\ , which means that
the solution $\hsigma$ has the same form as the static GMS soliton 
solution in the potential
\begin{equation}
U(\hsigma)=V(\frac 12\hsigma^2)-\frac 12\omega^2\hsigma^2\ .                                    
\end{equation}
Consider spherically symmetric solution~\cite{GMS} expanded
in terms of the projection operators,
\begin{equation}
\hsigma(\hat{x})=\sum_{n=0}^\infty\lambda_nP_n\ ,                                             \label{SigmaExp}
\end{equation}
where $P_n\equiv|n><n|$\ .
Replace $\hsigma$ in (\ref{NHam}), (\ref{N}), and the equation of motion
(\ref{NEqnM}),
\begin{eqnarray}
H&=&2\pi\sum_n[(n+1)(\lambda_{n+1}-\lambda_{n})^2+\theta
U(\lambda_n)]+\omega N\ ,\label{NewH}\\
N&=&2\pi\theta\omega\sum_n\lambda_n^2\ ,\label{NewN}
\end{eqnarray}
\begin{equation}
(n+1)(\lambda_{n+1}-\lambda_n)-n(\lambda_n-\lambda_{n-1})=\frac \theta
2 U(\lambda_n)\ .\label{NewEqnM}
\end{equation}
Sum the equation of motion from $n=0$ to $n=K$
\begin{equation}
\lambda_{K+1}-\lambda_K=\frac\theta{2(K+1)}\sum_{n=0}^K U'(\lambda_n)\
, \label{EqnSum}
\end{equation}
where $K\geq 0$ is an arbitrary integer.  A particular set of
$\lambda_n$'s defines a solution. Many properties of the solution can
still be derived from Eqn. (\ref{NewH}-\ref{EqnSum}) though a closed
form has not been constructed.  For example, because of the
finiteness of both the energy $H$ and the particle number $N$\ , we have,
\begin{equation}
\lambda_{n+1}=\lambda_n\ ,\qquad \lambda_n=0\ ,~~~\mbox{for}~~
n\rightarrow \infty\ ,                                                                          \label{LambdaC}
\end{equation}

\subsection{$Q$-ball solutions}

In static GMS soliton theory, the global minimum of the potential is
generally assumed to be at the origin, and the core of the soliton is
localized at the local minimum of the potential~(false bubble
solution). It is the noncommutativity that forbids the classical
decay of the solitons.  The corresponding commutative potential does
not have nontrivial topological structures, and hence yields no
soliton solutions. Therefore NC GMS solitons are genuine
NC effects and they disappear at small enough $\theta$\ ,
where the commutative limit is approached.

This is not the case with $Q$-ball solutions. The existence and
stablity of $Q$-ball solution rely on the conservation of the
charge $Q$ as the consequence of the global symmetry. The potential
for $Q$-ball solutions does not have nontrivial topological
structure. Therefore NC $Q$-balls are expected to exist even for very
small $\theta$ . We'll show that such NC $Q$-ball solutions would
smoothly reduce to the $Q$-ball solution in the commutative limit.

In the following we discuss the existence of NC $Q$-ball solutions in
a typical potential form, 
\begin{equation}
U(\sigma)=V(\sigma^2)-\frac
12\omega^2\sigma^2=a\sigma^2-b\sigma^4+c\sigma^6\ ,                                                \label{PotU}
\end{equation}
where the coefficients $b$ and $c$ are larger than zero, and $a=\frac
12(m^2-\omega^2)$\ . 

$U(\sigma)$ varies for different $\omega$\ . If
$\omega^2>m^2$ or $a<0$, $U(\sigma)$ has a local
maximum at the origin. In the commutative limit there is only a plane
wave solution.  Here similar plane wave solution in NC limit can also
be constructed.  Since for a stabe soliton solution $\lambda_n$ would have
to take values between $s$ and the origin and monotonically decrease in $n$
 \cite{Dur}, a simple argument can show that solitons cannot exist. 
There is a constraint that 
$\sum_{n=0}^K U'(\lambda_n)$
converges to zero as $K$ goes to infinity, which cannot be satisfied
in this case. To prove this constraint, suppose that
\begin{equation}
\sum_{n=0}^\infty U'(\lambda_n)\sim v\neq 0\ ,
\end{equation}
Sum Eqn. (\ref{EqnSum}) from a particular $K=q$ sufficiently large to
a point $p$ close to infinity,
\begin{equation}
\lambda_p-\lambda_q\sim \sum_{K=q}^{p}\frac vK\ .
\end{equation}
It's then easy to see $\lambda_p$ will not converge to zero as $p$
goes to infinity.

When $\omega^2<\nu^2$ , $\nu^2=m^2-{b^2}/{2c}$ , $U(\sigma)$
has only a global minimum at the origin. Even though in the
commutative theory no soliton solutions exist, for NC theory at
sufficiently large $\theta$\ , there are GMS type solitons exist. It has
been shown that there is a critical lower bound on $\theta$ for the
existence of NC soliton~\cite{Zhou}. Similar bounds would be expected
to exist for NC $Q$-ball for $0<\omega<\nu$ as well.

As $\nu^2<\omega^2<m^2$, $U(\sigma)$ has a local mininum at
the origin, a global minimum at $s$ ($U(s)<0$) and a zero 
$w=[(b-\sqrt{b^2-4ac})/2a]^{1/2}$ between $0$ and $s$ ($0<w<s$). In
the commutative case such potential form enables the existence of $Q$-
ball solutions. In the NC case it is expected that NC-$Q$ ball
solutions exist even for small $\theta$. In the following we take the
continuum limit of Eqn. (\ref{EqnSum}) for very small $\theta$ ,
and show that all the solitons $sP_K$ exist at $\theta\rightarrow
\infty$ converges to the commutative $Q$-ball solution as
$\theta\rightarrow 0$\ .

For very small $\theta$, all $\lambda_n$'s can be considered as
sufficiently close. Therefore $\lambda_n$ can be approximated by a
continuous function $\lambda(u)$~\footnote{Thanks to Dr. Shabnov for
helping on this point}. Let $u=K\theta$ , and
$\lambda_K=\lambda(K\theta)=\lambda(u)$ . Eqn. (\ref{EqnSum}) becomes
\begin{equation}
\lambda'(u)=\frac 1{2(u+\theta)}\int_0^{u}U'[\lambda(s)]ds+{\cal
O}(\theta)\ .
\end{equation}
Ignore ${\cal O}(\theta)$ term, we have
\begin{equation}
\frac{d\lambda}{du}+u\frac{d^2\lambda}{du^2}=\frac
12\frac{dU}{d\lambda}\ .
\end{equation}
Let $u=\frac 12v^2$ ,
\begin{equation}
  \frac{d^2\lambda}{dv^2}+\frac 1v\frac{d\lambda}{dv}=\frac
  {dU(\lambda)}{d\lambda}\ .
\end{equation}
This is exactly the equation of motion for the commutative $Q$-ball 
solution $\lambda(v)$ , 
with $v$ identified as radius $r$ . This can be explained as the following. The
Weyl transform of $\frac 12r^2$ is $a^\dag
a$ , and $a^\dag a$ has the eigenvalue $n\theta$ on the state $|n>$ .
As $\theta$ gets smaller, the eigenvalues $n\theta$ gets closer, and eventually
becomes continuous as $\frac 12r^2$ in the commuative limit. The coefficient
$\lambda_n$ just becomes the field $\lambda(r)$ in this limit.
In this description 
the commutative $Q$-ball can be considered as the analytical continuation of 
the NC $Q$-balls in $\theta$ .  

The formula for the energy and particle number in the commutative
limit can also be recovered by taking the continuum limit of
Eqn. (\ref{NewH}) and (\ref{NewN})\ ,
\begin{eqnarray}
  H&=&2\pi\int_0^\infty du \left[u(\frac
  {d\lambda}{du})^2+U(\lambda)+\omega N\right]= 2\pi\int_0^\infty vdv\left[\frac 12(\frac
  {d\lambda}{dv})^2+U(\lambda)+\omega N\right]\ ,\\ 
  N&=&2\pi\omega\int_0^\infty du\lambda^2(u)=2\pi\omega\int_0^\infty vdv\lambda^2(v)\ .
\end{eqnarray}

The existence of the commutative $Q$-ball solution are
proved by considering an analogous problem in which a classical
particle moves in the one-dimensional potential
$-U(\lambda)$~\cite{Coleman}.  The field configuration $\lambda(v)$ of
the $Q$-ball starts from a unique value $\rho=\lambda(0)$ between the 
zero $w$ and the global minimum $s$, then monotonically decreases in $v$ , 
and approaches $0$ when $v\rightarrow \infty$\ . This property of
$\lambda(v)$ is consistent with those of the general stable NC 
soliton solutions $\lambda_n$ at finite $\theta$ . In~\cite{Dur}, it's 
found that there exist smooth $\theta$ families of spherically
symmetric solutions in which $\lambda_n$ is monotonically decreasing
in $n$ . In the infinite $\theta$ limit such solution is just
$sP_K$ . As $\theta$ decrease, $\lambda_0,\cdots,\lambda_K$
decrease from $s$ , while other $\lambda_n(n>K)$ starts to move away
from the origin towards $s$ , but the whole $\lambda_n$ series remain
monotonically decrease in $n$ . Since in the commutative limit $\lambda_n$ just 
becomes $\lambda(v)$ , one can conclude that as $\theta$ decreases from $\infty$ to
zero, $\lambda_0$ will decrease from $s$ and eventually to $\rho$ at the 
commutative limit. 

\subsection{Virial Relation}

The Hamiltonian (\ref{NHam}) in the function formalism is 
\begin{equation}
  H[\sigma]=2\pi\int rdr\left(\frac 12(\partial_i\sigma)^2+U(\theta\ , \sigma)+\omega N\right)\ ,
\end{equation}
where the potential $U$ has explicit dependence on $\theta$ 
through star product. Suppose $\sigma(x)$ is the $Q$-ball solution, 
$H[\sigma(x/a)]$ must be stationary at $a=1$\ . A change of the integration variables shows that 
\begin{equation}
  H[\sigma(x/a)]=2\pi\int rdr\left(\frac 12(\partial_i\sigma(x))^2+
a^2U(\frac{\theta}{a^2}\ ,\sigma(x))+\omega N\right)\ ,
\end{equation}
and
\begin{equation}
  \left.\frac {d}{da}H[\sigma(x/a)]\right|_{a=1}=2\pi\int rdr\left(2U(\theta\ , \sigma)
  -2\frac{\partial}{\partial\theta}U(\theta\ , \sigma)\right)=0\ .                                       \label{Vir}
 \end{equation}
Unlike the Virial theorem for $d=1$ space dimension, here the kinetic energy is scale invariant. 
Scaling dependence of the energy includes two seperate terms from the potential and from 
its dependence on $\theta$ through the star products. 
The significance of Eqn. (\ref{Vir}) is more explicit in GMS soliton case, where 
the potential energy 
\begin{equation}
2\pi\int rdrU(\sigma)=\theta\sum_{n=0}^{\infty}U(\lambda_n)>0\ . 
\end{equation}
The scaling variable $a$ can be thought of as the size of the NC soliton. 
While the positive potential energy favors shrinking of the soliton, but 
the NC star products keep it from decay. 

\section{Quantization of noncommutative $Q$-ball}

Solitons are extended objects exist in field theory, the properties of which receive 
quantum corrections as the fields are quantized. In this section 
we follow very closely to the canonical quantization procedure 
proposed in~\cite{ChristLee,LeeReport}. 
Then we evaluate the ultraviolet divergences in the quantum corrections to 
the soliton energy at very small $\theta$ . 

\subsection{Canonical Quantization}

The general procedure to investigate the properties of the solitons is
to expand the fields around the classical solution. Because the momentum
and particle number are conserved in the system, we'll 
have to impose the corresponding constraints to erase the zero-frequency 
modes in the expansion. 

We start by making a point canonical transformation of $\phi(x)$ , 
\begin{equation}
  \phi=\frac 1{\sqrt{2}}e^{-i\beta(t)}[\sigma(x-X(t))+\chi(x-X(t),t)]\
  ,                                                                                                \label{trans}
\end{equation}
\begin{equation}
  \chi(x-X(t),t)\equiv\chi_R(x-X(t),t)+i\chi_I(x-X(t),t)\ ,
\end{equation}
where $\beta(t)$ and $X^i(t)$ are the collective coordinates represent
the over-all phase and the center of mass position. Impose the constraints 
on $\chi$ to ensure
the above transformation is a canonical transformation with equal
number of degrees of freedom before and after,
\begin{equation}
  \int\sigma\chi_I=0\ ,~~~~~\int\chi_R\partial_i\sigma=0\ ,                                        \label{Const}
\end{equation}
where $i=1,2$ . The integral sign denotes two dimensional
integrations over $x$. The star product is suppressed.  Unless
indicated otherwise, from now on the differential $d^2x$ and the NC
star product are implied wherever applicable.
The above constraints also remove the perturbative zero mode
solutions in the meson field $\chi$\ .  Let
\begin{eqnarray}
  \chi_R(x,t)&=&\sum_{a=3}^{\infty}q_{Ra}(t)f_a(x)\ ,\\
  \chi_I(x,t)&=&\sum_{\da=2}^{\infty}q_{I\da}(t)g_\da(x)\ ,
\end{eqnarray}
where $f_a(x)$ and $g_\da(x)$ are the real normal functions satisfy 
\begin{eqnarray}
\int f_af_b&=&\delta_{ab}\ ,\\
\int g_{\da} g_{\db}&=&\delta_{\da\db}\ ,\\
\end{eqnarray}
and  under the constraints, 
\begin{eqnarray}
  \int \partial_i\sigma f_a&=&0\ ,\\
  \int \sigma g_\da&=&0\ ,
\end{eqnarray}
where $a=3,4,\ldots$ and $\da=2,3,\ldots$ always in this paper. 

Rewrite the Lagrangian (\ref{action}) with (\ref{trans}),
\begin{equation}
  L=\frac 12 \dot{q}^T{\cal M}\dot{q}+{\cal V}(q),                                                      \label{CollLag}
\end{equation}
where $q^T=(X_1\ ,X_2\ ,\beta\ ,q_{R3}\ ,\ldots\ ,q_{I2}\ ,\ldots)$ and $T$ denotes 
matrix transpose, and 
\begin{equation}
  {\cal V}(q)\equiv\int(\partial_i\bar{\phi}\partial_i\phi+V(\frac 12\{\bar{\phi}\ ,\phi\}))            \label{CalPot}
\end{equation}                                                                                         
The matrix
elements of the symmetric ${\cal M}$ are
\begin{eqnarray}
{\cal M}_{ij}&=&M_0\delta_{ij}+\int\left(2\partial_i\sigma\partial_j\chi_R+\partial_i\chi_R
\partial_j\chi_R+\partial_i\chi_I\partial_j\chi_I\right)\ ,\\ 
{\cal M}_{\beta\beta}&=&I+\int(2\sigma\chi_R+\chi_R^2+\chi_I^2)\ ,\\ 
{\cal M}_{\beta i}&=&\int\left(-2\partial_i\sigma\chi_I+\chi_R\partial_i\chi_I-\chi_I\partial_i\chi_R\right)\
,\\ 
{\cal M}_{ia}&=&-\int f_a\partial_i\chi_R\ ,\\
{\cal M}_{i\da}&=&-\int g_{\da} \partial_i\chi_I\ ,\\
{\cal M}_{\beta a}&=&\int f_a\chi_I\ ,\\
{\cal M}_{\beta \da}&=&-\int g_{\da}\chi_R\ ,\\ 
{\cal M}_{ab}&=&\delta_{ab}\ ,\\
{\cal M}_{\da\db}&=&\delta_{{\da}{\db}}\ .
\end{eqnarray}
where $M_0\equiv\frac 12\int(\partial_i\sigma)^2$ and
$I\equiv\int\sigma^2$\ .
The conjugate momentum of $q$ is
\begin{equation}
  p=M\dot{q}\equiv(P_1\ ,P_2\ ,N\ ,p_{R3}\ ,\ldots\ ,p_{I2}\ ,\ldots)^T\ .
\end{equation}
The particle number $N$ and the total momentum $P_i$ are conserved 
since the Lagrangian (\ref{CollLag}) is independent of the collective coordinates 
$\beta$ and $X^i$. 
Quantize the new canonical coordinates, 
\begin{eqnarray}
  [X^i\ ,P^j]&=&i\delta^{ij}\ ,\\
  ~[\beta\ ,N]&=&i\ ,\\
  ~[q_{Ra}\ ,p_{Rb}]&=&i\delta_{ab}\ ,\\
  ~[q_{I\da}\ ,p_{I\db}]&=&i\delta_{\da\db}\ .
\end{eqnarray}
The Hamiltonian 
\begin{equation}
  H=\frac 12J^{-1}p^T{\cal M}^{-1}Jp+{\cal V}(q)\ ,
\end{equation} 
where $J=\sqrt{\det {\cal M}}$ because the operator ordering in $H$ is unambiguously 
determined by the 
ordinary quantized Hamiltonian with the coordinates $\phi$~\cite{ChristLee}. 

The quantum states can be labelled as $|P^1\ ,P^2\ ,N\ ,q_{Ra}\ ,q_{I\da}>$\ . One can  
solve the Schr\"{o}dinger Equation perturbatively around the one soliton ground state 
$|P^1=P^2=0\ ,N=I\omega\ ,0>$\ . In this state $P^i$ and $N$ are the momentum and particle 
number of the classical solution $\sigma$, which can be obtained by letting $\phi=\sigma$ in 
Eqn. (\ref{NewN}) and $P^i=\int\dot{\bar{\phi}}\partial^i\phi+\partial^i\bar{\phi}\dot{\phi}$  
\cite{Xiong}. $0$ labels the lowest energy state with the given $P^i$ and $N$ value. 

We can then treat $\chi_R$ and $\chi_I$ as 
perturbative degrees of freedom, and expand the Hamiltonian perturbatively around the 
one soliton ground state order by  
order in the weak coupling constant $g$ , defined in Eqn. (\ref{ScalePot}).

$\sigma$ is at the order of $g^{-1}$ as a soliton solution. $M_0$ and $I$ are 
$g^{-2}$ order. Then $P_i$ and $N$ are at the $g^{-2}$ order, while $p_{Ra}$ and 
$p_{I\da}$ at $g^0$ order. 
Since $J$ commute with $P_i$ and $N$ , and at the leading $g^{-3}$ order, 
$J=M_0\sqrt{I}$ is a constant or $[p\ ,J]=0$\ , one can check that   
$J$ would not be a factor in the Hamiltonian up to the order $g^0$\ . 

The Hamiltonian can be expanded order by order, $H=H_0+H_1+H_2$\ ,
with the expansion relation, 
\begin{equation}
 {\cal M}^{-1}={\cal M}_0^{-1}+{\cal M}_0^{-1}\Delta{\cal M}_0^{-1}+
 {\cal M}_0^{-1}\Delta{\cal M}_0^{-1}\Delta{\cal M}_0^{-1}+\cdots\ , 
\end{equation}
where ${\cal M}={\cal M}_0+\Delta$, and ${\cal M}_0$ has only nonzero diagonal   
elements, ${\cal M}_0^{qq}=(M_0\ ,M_0\ ,I\ ,1\ ,1)$\ .  

$H_0$\ , equal to the energy of the
classical solution, is at the order of $g^{-2}$\ , 
\begin{equation}
  H_0=M_0+\frac 12I\omega^2+V(\frac 12\sigma^2)\ ,
\end{equation}

$H_1$\ , linear in $\chi$\ , vanishes due to the fixed $N$ and $P_i$ , 
which ensures that $\chi_R$ and $\chi_I$ are at the order of $g^0$\ .  

The term quadratic in $\chi$ is at the $g^0$ order, 
\begin{equation}
  H_2=\frac 12(p_{Ra}-\omega\int f_a\chi_I)^2+
\frac 12(p_{I\da}+\omega\int g_\da\chi_R)^2
+2\frac{\omega^2}{M_0} (\int\partial_i\sigma\chi_I)^2+2\frac
{\omega^2}{I}(\int\sigma\chi_R)^2+{\cal V}_2(q)\ ,                                                 \label{MesonHam}
\end{equation}                                                                                     
where
\begin{eqnarray}
  {\cal V}_2(q)&=&\int\left\{\frac 12[(\partial_i\chi_R)^2+
(\partial_i\chi_I)^2]-\frac 12\omega^2(\chi_R^2+\chi_I^2)\right.\nonumber\\ 
&&\left. +\frac
12(\chi_R^2+\chi_I^2)V'(1/2\sigma^2)
+V\left(\frac 12\sigma^2\ ,\frac 12\{\chi_R, \sigma\}\ ,
\frac 12\{\chi_R, \sigma\}\right)\right\}\ ,                                                      \label{MesonPot}
\end{eqnarray}
where 
$V\left(\frac 12\sigma^2\ ,\frac 12\{\chi_R, \sigma\}\ ,\frac
12\{\chi_R, \sigma\}\right)$ represents the terms from the expansion of the 
potential $V$ quadratic in $\chi_R$\ .

\subsection{Energy Corrections at Very Small $\theta$}

The Hamiltonian is seperated into two parts, described by the baryon 
degrees of freedom ($P^i\ ,N$) and meson degrees of freedom ($\chi_R\ ,\chi_I$) respectively. 
The sum of the frequencies of the meson excitations is the zero-point energy of 
$H_2$, 
\begin{equation}
   <P^1=P^2=0\ ,N=I\omega\ ,0|~H_2~|P^1=P^2=0\ ,N=I\omega\ ,0>\ ,                                    \label{ZeroPoint}
\end{equation}
which, subtracted by the vacuum energy 
$E_{\hbox{vac}}=\int {d^2k}/({2\pi})^2\sqrt{k^2+m^2}$\ , gives the quantum corrections 
to the soliton energy. 

${\cal V}_2(q)$ is the perturbative expansion of the effective potential 
${\cal V}(q)-\omega Q[\phi,\bar{\phi}]$\ , Eqn. (\ref{CalPot}) and (\ref{ConCharge}), 
around the solution $\sigma$ . 
It's easy to check that $\chi_R=\partial_i\sigma$ and $\chi_I=\sigma$ are the eigenmodes of 
${\cal V}_2(q)$ with eigenfrequency 0, due to the translational and rotational 
invariance of the potential. Therefore we can 
define the normal functions $f_a$ and $g_\da$ to be the eigenmodes of ${\cal V}_2$ , or  
\begin{equation}
  {\cal V}_2(q)=\frac 12\Omega_{Ra}^2q_{Ra}^2+\frac 12\Omega_{I\da}^2q_{I\da}^2\ , 
\end{equation}
with the frequencies $\Omega_{Ra}$ and $\Omega_{I\da}$ . The potential ${\cal V}_2$ 
are highly nonlocal since the fields are multiplied by NC star product. 
In the commutative $Q$-ball case, ${\cal V}_2$ has been shown to have only one s-wave 
eigenmode in $\chi_R$ sector with imaginary frequency $\Omega_{R3}$~\cite{LeeReport}. 
In last section, we have shown that as NC parameter $\theta$ is taken to be small enough, 
the NC soliton solution will reduce arbitrary close to its commutative analog. 
Therefore close to the commutative limit ${\cal V}_2$ is 
expected to have the similar eigenvalues and eigenmodes as its commutative 
analog. NC $Q$-ball is also expected to be stable as its commutative analog. 
We'll assume $\theta$ is chosen to be such a small value in evaluating 
the quantum effects of the noncommutativity. 

Define $f_i=1/\sqrt{M_0}\partial_i\sigma$ and $g_1=1/\sqrt{I}\sigma$ , 
Rewrite the Hamiltonian $H_2$ , (\ref{MesonHam}), in the matrix form, 
\begin{equation}
  H_2=\frac 12(\cP^T-\omega\cQ^T\Upsilon^T)(\cP-\omega\Upsilon\cQ)
      +2\omega^2\cQ^T\Xi\cQ+\frac 12\cQ^T\Omega^2\cQ\ ,                                         \label{MatrixHam}
\end{equation}
where the matrices are defined in the following,
\begin{equation}
  \cP^T\equiv (p_{Ra}\ ,p_{I\da})\ ,\qquad 
  \cQ^T\equiv (q_{Ra}\ ,q_{I\da})\ ,
\end{equation}
\begin{equation}
  \Upsilon\equiv\left(\matrix{ 0 & \Gamma_{a\da} \cr \\
  -\Gamma_{\da a}^T & 0 }\right)\ ,\qquad
  \Xi\equiv\left(\matrix{ {\cal F}_{ab} & 0 \cr \\
    0 & {\cal G}_{\da\db} }\right)\ ,\qquad 
  \Omega\equiv\left(\matrix{ \Omega_{Ra} & 0 \cr \\
  0 & \Omega_{I\da} }\right)\ ,
\end{equation}
where 
\begin{equation}
  \Gamma_{a\da}\equiv\int f_ag_\da\ ,\qquad
  {\cal F}_{ab}\equiv\int g_1f_a\int g_1f_b\ ,\qquad
  {\cal G}_{\da\db}\equiv\int f_ig_\da\int f_ig_\db\ .
\end{equation}
The equation of motion, 
\begin{equation}
  \dot{\cQ}=\frac{\partial H_2}{\partial \cP}\ ,\qquad 
\dot{\cP}=-\frac{\partial H_2}{\partial \cQ}\ ,
\end{equation}
give
\begin{eqnarray}
  \dot{\cQ}&=&\cP-\omega\Upsilon\cQ\ , \\                                                         \label{cP}
  \dot{\cP}&=&\omega\Upsilon^T(\cP-\Upsilon\cQ)-4\omega^2\Xi\cQ-\Omega^2\cQ\ .
\end{eqnarray}
Therefore, 
\begin{equation}
  \ddot{\cQ}+2\omega\Upsilon\dot{\cQ}+4\omega^2\Xi\cQ+\Omega^2\cQ=0\ .
\end{equation}
Let the real normal eigenmodes of $\cQ$ be 
\begin{equation}
\cQ^A=(\xi_{Ra}^A\ ,\xi_{I\da}^A)^T, 
\end{equation}
where ${\cQ^A}^T\cQ^B=\delta^{AB}$\ .
Replace $\cQ=\cQ^A\hbox{exp}(-i\Lambda_At)$ (Index $A$ is not summed over) 
in the above equation. Since ${\cQ^A}^T\Upsilon\cQ^A=0$ , 
\begin{equation}
\Lambda_A=\sqrt{{\cQ^A}^T(4\omega^2\Xi+\Omega^2)\cQ^A}\ .                                         \label{EigenFreq}
\end{equation}
Introduce creation and annihilation operators, $[{\cal C}_A\ ,{\cal C}_B^\dag]=\delta_{AB}$\ ,
$\cQ$ can then be quantized as, 
\begin{equation}
  \cQ=\sum_A\frac {\cQ^A}{\sqrt{2\Lambda_A}}
({\cal C}_Ae^{-i\Lambda_At}+{\cal C}_A^\dag e^{i\Lambda_At})\ . 
\end{equation}
Use this equation and Eqn. (\ref{cP}) and (\ref{MatrixHam}), one can 
define the one soliton ground state, 
\begin{equation}
  {\cal C}_A|P^1=P^2=0\ ,N=I\omega\ ,0>=0\ ,
\end{equation}
then the zero-point energy of $H_2$ (\ref{ZeroPoint}) is 
\begin{equation}
  \frac 12\sum_A\Lambda_A=\frac 12\Tr\{\Lambda\}\ ,                                          \label{ZeroPoint2}
\end{equation}
where the matrix $\Lambda$ is diagonal with the eigenvalues $\Lambda_A$\ .

In the commutative theory the zero-point energy contains the divergences
even after subtraction of the vacuum energy. The finiteness of the soliton energy 
is recovered by starting from the renormalized form of the action (\ref{action}), 
which induces the counter terms also contain the divergences~\cite{Rajaraman}. 

Work in the specific form of the $\phi^6$ potential (\ref{PotU}), 
\begin{equation}
  V(\frac 12\{\bar{\phi},\phi\})=m^2(\frac 12\{\bar{\phi},\phi\})                              
-bm^2g^2(\frac 12\{\bar{\phi},\phi\})^2+cm^2g^4(\frac 12\{\bar{\phi},\phi\})^3\ .                   \label{SpecV}
\end{equation}
At the $g^0$ order, or the one-loop order, 
the general formula for the soliton energy is 
\begin{eqnarray}
  E_{\hbox{soliton}}&\equiv&<P^1=P^2=0\ ,N=I\omega\ ,0|~H~|P^1=P^2=0\ ,N=I\omega\ ,0>\\
           &=&H_0+\frac 12\hbox{Tr}\{\Lambda\}-E_{\hbox{vac}}+   
\frac 12\delta m^2\int\sigma^2-bm^2\delta g^2_{(4)}\int\sigma^4\ ,                                  \label{SolEnergy}
\end{eqnarray}
where $\delta m^2$ and $\delta g^2_{(4)}$ are the counter terms for the mass and the 
$\phi^4$ coupling respectively. The $\phi^6$ coupling doesn't receive 
loop corrections. The $\phi^4$ coupling terms yield the right coefficients and can 
be renormalized~\cite{Kos}. 

The loop integration in the NC field theory generally contains phase factors 
which yield the 
interesting UV-IR phenomenon upon renormalization~\cite{MRS}. In the following 
we evaluate the quantum correction from the zero-point energy of 
$H_2$ in Eqn. (\ref{ZeroPoint2}) 
and show that it contains the same phase factors as those appears in 
the counter terms $\delta m^2$ and $\delta g^2_{(4)}$\ . 

We start by arguing that only $1/2\hbox{Tr}\{\Omega\}$ is needed in evaluating the 
leading divergence. 
In Eqn. (\ref{EigenFreq}), it's easy to see ${\cQ^A}^T\Xi\cQ^A$ is finite, 
\begin{eqnarray}
  {\cQ^A}^T\Xi\cQ^A&=&(\int g_1f_a\xi_{Ra})^2+(\int f_ig_\da\xi_{I\da})^2 \nonumber \\
                 &\leq&[\int g_1^2+\int (f_a\xi_{Ra})^2]^2+[\int f_1^2+\int (g_\da\xi_{I\da})^2]^2
                  +[\int f_2^2+\int (g_\da\xi_{I\da})^2]^2    \nonumber \\
                 &=&(1+\xi_{Ra}^2)^2+2(1+\xi_{I\da}^2)^2~\leq~ 12\ .
\end{eqnarray}
As we'll see that the eigenvalues $\Omega_{Ra}$ and $\Omega_{I\da}$ 
behave like $\sqrt{k^2-\omega^2+m^2}$ at very large $k$\ .
The leading divergence of 
$\hbox{Tr}\{\Lambda\}$ will be determined by $\Tr\{\Omega\}$\ . 

$\Omega_{Ra}$ and $\Omega_{I\da}$\ , eigenfrequencies of ${\cal V}_2(q)$ 
in Eqn. (\ref{MesonPot}), satisfy the linear equations, 
\begin{eqnarray}
  &&(-\partial_i^2-\omega^2+m^2)\chi_I-\frac 12bm^2g^2\{\sigma^2,\chi_I\}
+\frac 38cm^2g^4\{\sigma^4,\chi_I\}=\Omega_{I\da}^2\chi_I\ ,                                     \label{ChiIPot}
\\  &&(-\partial_i^2-\omega^2+m^2)\chi_R
-bm^2g^2(\{\sigma^2,\chi_R\}+\sigma\chi_R\sigma)+ \nonumber\\
&&\qquad\qquad\qquad\frac 34cm^2g^4(\{\sigma^4,\chi_R\}+\sigma^2\chi_R\sigma^2+\sigma\{\sigma,\chi_R\}\sigma)
=\Omega_{Ra}^2\chi_R\ .                                                                             \label{ChiRPot}
\end{eqnarray}

The above equations are just time independent Schr\"{o}dinger equations. 
In particular phase shifts from the central 
potential have been used in calculating the soliton energy correction~\cite{Graham}. 
The basic idea is that in the central potential for each partial wave, 
the difference of the density of the states between the scattered wave 
and the free wave is related to the derivative of the phase shift, 
\begin{equation}
  \rho_l(k)-\rho(0)=\frac 1\pi\frac {d\delta_l(k)}{dk}\ ,                                       \label{PhaseShift}
\end{equation}
where $l$ goes from $-\infty$ to $\infty$\ .
The finiteness of the particle number, $N=\omega\int\sigma^2$ , 
determines that $\sigma\rightarrow o$ as $r\rightarrow \infty$ . 
Therefore the NC potential in Eqn. 
(\ref{ChiIPot}) and (\ref{ChiRPot}) is radial symmetric and vanishes at 
$\infty$\ . For the most general potential term $\cW_F(r)\star\chi\star\cW_B(r)$ ,
\begin{equation}
      [\cW_F(r)\star\chi\star\cW_B(r)\ ,L]=\cW_F(r)\star[\chi\ ,L]\star\cW_B(r)\ , 
\end{equation}
where $L=-i\epsilon^{ij}x^i\partial^j$ is the angular momentum.
The star product is made explicit here and in the rest of the section. 
This formula can be easily proved in the Weyl transforms of the fields. 
Going to the momentum space, one can generalize the result in \cite{Big} and show that 
\begin{equation}
  \cW_F(x)\star\chi(x)\star\cW_B(x)=\int \frac{d^2p_f}{(2\pi)^2}\frac{d^2p_b}{(2\pi)^2}
\widetilde{\cW}_F(p_f)\widetilde{\cW}_B(p_b)
e^{i\bp_f(\bx+\fitheta\tbpartial)}e^{i\bp_b(\bx-\fitheta\tbpartial)}\chi(x)\ ,                             \label{EffW}
\end{equation}
where $\tpartial^i\equiv \epsilon^{ij}\partial^j$\ . 

Using Eqn. (\ref{PhaseShift}), consider only the leading divergence, 
we have~\cite{Sch}, 
\begin{eqnarray}
  \frac 12\Tr\{\Lambda\}-E_\hvac&\sim&\frac 12\Tr\{\Omega\}-E_\hvac\nonumber\\
&\sim&\frac 1{2\pi}\int d\sqrt{k^2+m^2}\sum_l[\delta_{Il}(k)+\delta_{Rl}(k)]\ ,                 \label{EnergyDiv}
\end{eqnarray}
where $\delta_{Il}(k)$ and $\delta_{Rl}(k)$ are the phase shifts for 
$\chi_I$ and $\chi_R$\ . 
The sum of the phase shifts can be evaluated through Born approximation. 
In the commutative case, this leads to the cancellation of the tadpole 
diagram~\cite{Graham}. 

Eqn. (\ref{ChiIPot}) and (\ref{ChiRPot}) have the Jost solution form at large $r$ , 
\begin{equation}
  \chi\sim h_l^*(kr)+e^{2i\delta_l}h_l(kr)\ .
\end{equation}
Considering the asymptotic ($r\rightarrow\infty$) behaviour of the solution, 
the standard procedure \cite{Sakurai} leads to the scattering amplitude, 
\begin{equation}
  f(\bk ',\bk)=f(\phi)=\sum_lf_l(k)e^{il\phi}
=\frac 1{\sqrt{k}}\sum_le^{i\delta_l}\sin\delta_le^{il\phi}\ ,
\end{equation}
where $k'=k$ and $\phi$ is the angle between $\bk '$ and $\bk$\ . 
At large $l$\ , or $\delta_l\approx 0$\ , we can see 
\begin{equation}
  \sum_l\delta_l\approx \sqrt{k}f(\phi=0) 
\end{equation}

$f(\bk ',\bk)$ can also be calculated through Born approximation, 
replacing $\chi$ by $e^{-i\bk\bx}$ in the potential form (\ref{EffW}), 
\begin{eqnarray}
  f(\bk ',\bk)&=&-\frac 1{4\sqrt{k}}\int d^2xe^{-i\bk ' \bx}
\sum_\rmi\cW_F^{(\rmi)}\star e^{-i\bk \bx}
\star\cW_B^{(\rmi)}\nonumber\\
&=&-\frac 1{4\sqrt{k}}\int d^2xe^{-i(\bk '-\bk)\bx}\sum_\rmi
\cW_F^{(\rmi)}(x-\ftheta \tk)\cW_B^{(\rmi)}(x+\ftheta \tk)
\end{eqnarray}
where ${\rm i}$ labels the potential terms in Eqn. (\ref{ChiIPot}) and 
(\ref{ChiRPot}), and $\tk^i\equiv\epsilon^{ij}k^j$\ . Therefore 
\begin{eqnarray}
  \sum_l\delta_l&=&-\frac 14\int d^2x\sum_\rmi\cW_F^{(\rmi)}(x-\ftheta\tk)
\cW_B^{(\rmi)}(x+\ftheta\tk)\nonumber\\
&=&-\frac 14\int \frac{d^2p}{(2\pi)^2}\sum_\rmi\widetilde{\cW}^{(\rmi)}_F(p)
\widetilde{\cW}^{(\rmi)}_B(-p)e^{-i\theta \bp\tbk}\ .
\end{eqnarray}
The right hand side only depends on the magnitude $k$ due to the central 
potential $\cW(r)$. 

Now we are ready to evaluate Eqn. (\ref{EnergyDiv}), 
\begin{eqnarray}
  &&\frac 1{2\pi}\int d\sqrt{k^2+m^2}\sum_l[\delta_{Il}(k)+\delta_{Rl}(k)] \nonumber \\
  &=&-\frac 1{8\pi}\int  d\sqrt{k^2+m^2} 
\int \frac{d^2p}{(2\pi)^2}\sum_\rmi\widetilde{\cW}^{(\rmi)}_F(p)
\widetilde{\cW}^{(\rmi)}_B(-p)e^{-i\theta \bp\tbk}\nonumber\\
&=&-\frac 12\int \frac{d^2p}{(2\pi)^2}\int \frac{d^2k}{(2\pi)^2}
\frac{e^{-i\theta \bp\tbk}}{2\sqrt{k^2+m^2}}
\sum_\rmi\widetilde{\cW}^{(\rmi)}_F(p)\widetilde{\cW}^{(\rmi)}_B(-p)\ .                            \label{ZeroPoint3}
\end{eqnarray}
The integration over k is exactly part of the tadpole diagram belongs to $\delta m^2$
~\cite{Kos}, and it contains the UV/IR divergence ($\Lambda\rightarrow \infty$ and 
$p\rightarrow 0$) evaluated with the cutoff $\Lambda$~\cite{MRS}, 
\begin{equation}
  \int \frac{d^2k}{(2\pi)^2}\frac{e^{-i\theta \bp\tbk}}{2\sqrt{k^2+m^2}}=
\frac 2{(4\pi)^{3/2}}(m\Lambda_{p})^{1/2}K_{\frac 12}\left(\frac{2m}
{\Lambda_{p}}\right)=\frac{\Lambda_{p}}{8\pi}+{\cal O}(1)\ ,                                     \label{UVIR}
\end{equation}
where $\Lambda_{p}\equiv(\theta^2p^2/4+1/\Lambda^2)^{-1/2}$\ .
Notice that the above UV/IR divergence from Eqn. (\ref{ZeroPoint3}) occurs 
only when both $\cW_F$ and $\cW_B$ exist. In other words, only the 
terms $\sigma\chi_R\sigma\ ,\sigma^2\chi_R\sigma^2$ and $\sigma\{\sigma,\chi_R\}\sigma$ 
in Eqn. (\ref{ChiIPot}) and (\ref{ChiRPot}) yield UV/IR divergence. 
All other potential terms only give the normal UV divergence where the phase 
factor is absent. Since the counter term $\delta m^2$ and $\delta g^2$ do not 
include UV/IR divergence, we are certain that the $Q$-ball energy 
correction includes UV/IR 
divergence. Cancellation of the UV divergences is not obvious 
because the exact value of the eigenfrequencies $\Lambda_A$ is unknown. 

\section{Finite $\theta$ and Noncommutative GMS Solitons}
The above calculation assumes that the NC parameter $\theta$ is sufficiently 
small so that the NC potential will generate the Jost solution form as 
in the commutative case. 
Let's consider the effects of the NC potential (\ref{EffW}) 
in the case that $\theta$ is not small. 

Since 
\begin{equation}
   [x^i\pm\fitheta\tpartial^i,x^j\pm\fitheta\tpartial^j]=\pm
i\theta\epsilon^{ij}\ ,
\end{equation} 
the effective scattering potential for the NC 
interaction $\cW_F(x)\star\chi\star\cW_B(x)$ (\ref{EffW}) is just 
\begin{equation}
\widehat{\cW}_F(x+\fitheta\tpartial)
\widehat{\cW}_B(x-\fitheta\tpartial)\ ,
\end{equation}
multiplication of the Weyl transforms of 
$\cW_F(x)$ and $\cW_B(x)$\ . Notice $\widehat{\cW}_F$ and $\widehat{\cW}_B$ 
commute since $[x^i+i\theta/2\tpartial^i,x^j-i\theta/2\tpartial^j]=0$\ .

Now considering 
\begin{equation}
  \widehat{\cW}_F(x)=\int\frac{d^2p_f}{(2\pi)^2}\widetilde{\cW}_F(p_f)
e^{i\bp_f(\bx+\fitheta\tbpartial)}\ ,
\end{equation}
the noncommutativity, 
\begin{equation}
[p_f^1(x^1+\fitheta\tpartial^1)\ ,
p_f^2(x^2+\fitheta\tpartial^2)]=i\theta p_f^1p_f^2\ ,                                         \label{Noncom}
\end{equation}
can be suppressed 
even if $\theta$ is not small, as long as $\cW(x)$ is smooth enough 
or $\widetilde{\cW}(p_f)\rightarrow 0$ at large $p_f$\ . 
Notice small $p_f$ is also the IR limit 
we discussed in the last section. Under this assumption we can write
\begin{equation}
  \widehat{\cW}_F(x)\approx\cW_F(x+\fitheta\tpartial)=
\cW_F(|\bx+\fitheta\tbpartial|^2)=\cW_F(r^2-\ftheta L+\partial_i^2)\ .
\end{equation}
Acting on the field 
$\chi(x)=u_l(kr)e^{il\phi}$, the effective potential becomes 
$  \cW_F(r^2-{\theta l}/{2}+\partial_i^2)$\ . Similar calculation 
applied to $\widehat{\cW}_B(x)$ yields $\cW_B(r^2+{\theta l}/{2}+\partial_i^2)$\ . 
Therefore at large $k$ and 
large $r$, we can treat the scattering potential perturbatively as in the 
commutative case.    
The phase shift evaluation of the energy of the soliton could still 
apply provided that $\cW$ or the soliton solution $\sigma$ are smooth enough. 

NC GMS soliton only exists at finite $\theta$\ . Based on 
the above arguments, we can still evaluate its quantum corrections with 
the phase shift method in the last section. 

Quantization of GMS soliton and $Q$-ball share a lot of similarities.  
To get the GMS soliton theory, we make the replacements 
($\phi\ ,\bphi\rightarrow 1/{\sqrt{2}}\Phi$) in the previous
complex scalar field theory (\ref{action}). With the potential (\ref{SpecV}), 
the Lagrangian becomes,
\begin{equation}
  {\cal L}=-\frac 12(\partial_\mu\Phi)^2+V(\Phi)=-\frac 12(\partial_\mu\Phi)^2+\frac 12m^2\Phi^2
-\frac 14bm^2g^2\Phi^4+\frac 18cm^2g^4\Phi^6\ ,                                                     \label{PhiLag}
\end{equation}
where $\Phi$ is multiplied by the star product. The renormalizability of the theory is
proved in~\cite{Chepelev}. Let $\omega=0$ because there is no 
conserved charge $Q$ (\ref{ConCharge}) in the theory. When $b^2-4c<0$\ ,
the potential in (\ref{PhiLag}) has a local minimum at $\sqrt{b}g/2$ besides the global 
minimum at the origin, and the GMS soliton solution $\sigma$ exists. 
Replace the expansion (\ref{trans}) by $\Phi=\sigma+\chi$ , where $\chi=\chi_R$ is 
real. As a result, the meson degrees of freedom are only $\chi_R$ or $\chi$\ . 
Upon quantization, the soliton energy is still given by Eqn. (\ref{SolEnergy}).
Since $\omega=0$ and $\Lambda=\Omega$ are the exact eigenfrequencies of $H_2$ (\ref{MesonHam}),
we have 
\begin{equation}
  (-\partial_i^2+m^2)\chi_R
-bm^2g^2(\{\sigma^2,\chi_R\}+\sigma\chi_R\sigma)
+\frac 34cm^2g^4(\{\sigma^4,\chi_R\}+\sigma^2\chi_R\sigma^2+\sigma\{\sigma,\chi_R\}\sigma)
=\Lambda_{a}^2\chi_R\ . 
\end{equation}
Eqn. (\ref{EnergyDiv}) describes the exact ultraviolet divergences in $1/2\Tr\{\Lambda\}
-E_\hvac$\ . Therefore we are able to check the cancellation of the divergences in Eqn. 
(\ref{SolEnergy})\ . As mentioned in the previous section, in the above equation, 
the terms $\sigma\chi_R\sigma$\ , $\sigma^2\chi_R\sigma^2$ and 
$\sigma\{\sigma,\chi_R\}\sigma$ yield UV/IR divergences, while the rest terms yield UV 
divergence. A critical observation is that those terms yield UV/IR divergence 
have one to one correspondence with the contractions of the fields yield Nonplanar 
Feynman diagrams~\cite{MRS}, and those terms yield UV divergence correspond exactly 
to the planar digrams. We can just spare the details of counting the divergences. 
Since the counter terms $\delta m^2$ and $\delta g^2_{(4)}$ cancel exactly the 
UV divergence part, we conclude that the soliton energy (\ref{SolEnergy}) is    
UV finite, but includes all the UV/IR divergences. 

\section{Conclusion and Discussion}

In this paper we discuss the quantization of NC solitons in ($2+1$) dimensional 
scalar field theory. In particular, classical solutions and quantization of the 
NC $Q$-balls at very small $\theta$ are investigated in detail. Classically 
NC $Q$-balls reduce to the commutative $Q$-ball as $\theta$ goes to zero. 
Quantum mechanically, because loop integrations in the NC field thoery have different 
ultraviolet structure from those in the commutative theory, i. e. UV/IR mixing, 
quantum corrections to the NC soliton energy necessarily include the UV/IR 
divergent terms which cannot be renormalized away. The existence of such terms 
in the energy is demonstrated through the phase shift summation. The same method 
is futher generalized to NC GMS solitons which exist only at finite $\theta$\ . 
In the small momentum limit, or for the suffciently smooth soliton solutions, 
divergence structure of the soliton energy can be calculated exactly. In this case the 
energy is found to contain no UV divergence but all the UV/IR divergences. 
In~\cite{Miao} quantum corrections to the NC soliton energy are also calculated but 
at very large $\theta$, where no UV/IR divergence is found. We believe that's 
because at large $\theta$, the noncommutativity (\ref{Noncom}) is not small and 
cannot be ignored, and the potential term is the dominant term instead of a perturbative 
one. In this case the phase shift sum is not a good approximation to the energy 
correction. 

An interpretation to the UV/IR divergence is given in~\cite{RS2}, 
where new light degrees of freedom are introduced in the Wilsonian effective action. 
UV/IR divergence can be reproduced by integrating out those new degrees of freedom, 
which are then interpreted as closed string modes with channel duality.  
It will be interesting to further consider the NC solitons in the gauge theory, 
where they are interpreted as D-branes~\cite{MinUnstable,HarveyDbrane} and 
D-brane action is properly 
recovered. If such NC solitons are quantized one might be able to recover the effective 
interaction between D-branes and closed strings .

\vskip 3cm
\centerline{\bf Acknowledgments} 
\vskip 0.3cm
I would like to thank Prof. 
Pierre Ramond for the encouragement and support throughout the time I work on this  
paper. Also I am grateful to Prof. Zongan Qiu for helpful discussions. This 
work was supported by the Department of Energy under grant
DE-FG02-97ER41029.


\newpage
\begin{thebibliography}{Ref}
\bibitem{SW} N. Seiberg and E. Witten, {\sl JHEP} {\bf 09} 032 (1999), {\it hep-th/9908142}.
\bibitem{Connes} A. Connes, M. R. Douglas and A. Schwarz, {\sl JHEP} {\bf 02} 003 (1998), {\it hep-th/9711162}. 
\bibitem{GMS} R. Gopakumar, S. Minwalla and A. Strominger, {\sl JHEP} {\bf 05} 020 (2000), {\it hep-th/0003160}.
\bibitem{Nek} N. A. Nekrasov, ``Trieste lectures on solitons in noncommutative gauge theories,'' 
{\it hep-th/0011095}, and the references therein.
\bibitem{MinUnstable} M. Aganagic, R. Gopakumar, S. Minwalla and A. Strominger, {\sl JHEP} {\bf 04} 001 (2001), 
{\it hep-th/0009142}. 
\bibitem{HarveyDbrane} J. A. Harvey, P. Kraus, F. Larsen and E. J. Martinec, {\sl JHEP} {\bf 07} 042 (2000), 
{\it hep-th/0005031}. 
\bibitem{Derrick} G. H. Derrick, {\sl J. Math. Phys.} {\bf 5} 1252 (1964). 
\bibitem{Zachos} T. Curtright, T. Uematsu and C. Zachos, ``Generating all Wigner Functions,'' {\it hep-th/0011137}, and 
the references therein.
\bibitem{LeeReport} T. D. Lee and Y. Pang, {\sl Phys. Rept.} {\bf 221} 251 (1992). 
\bibitem{Coleman} S. R. Coleman, {\sl Nucl. Phys.} {\bf B 262} 263 (1985).
\bibitem{Dur} B. Durhuus, T. Jonsson and R. Nest, 
``The Existence and stability of noncommutative scalar solitons,'' {\it hep-th/0107121}.
\bibitem{MRS} S. Minwalla, M. V. Raamsdonk and N. Seiberg, {\sl JHEP} {\bf 02} 020 (2000), {\it hep-th/9912072}.
\bibitem{NCQball} Y. Kiem, C. Kim and Y. Kim, {\sl Phys. Lett.} {\bf B 507} 207 (2001), {\it hep-th/0102160}. 
\bibitem{Komba} J. A. Harvey, ``Komaba lectures on noncommutative solitons and D-branes,'' {\it hep-th/0102076}.
\bibitem{Dirac} P. A. M. Dirac, {\sl Rev. of Mod. Phys.} {\bf 21} 392 (1949).
\bibitem{Zhou} C. Zhou, ``Noncommutative scalar solitons at finite theta,'' {\it hep-th/0007255}.
\bibitem{ChristLee} N. H. Christ and T. D. Lee, {\sl Phys. Rev.} {\bf D12} 1606 (1975).
\bibitem{Graham} N. Graham, R. L. Jaffe and H. Weigel, {\sl Int. J. Mod. Phys.} {\bf A17} 846 (2002), 
{\it hep-th/0201148}.
\bibitem{Xiong} T. Pengpan and X. Xiong, {\sl Phys. Rev.} {\bf D63} 085012 (2001), {\it hep-th/0009070}.
\bibitem{Rajaraman} R. Rajaraman and E. J. Weinberg, {\sl Phys. Rev.} {\bf D11} 2950 (1975).
\bibitem{Sch} J. Schwinger, {\sl Phys. Rev.} {\bf 94} 1362 (1954).
\bibitem{Big} D. Bigatti and L. Susskind, {\sl Phys. Rev.} {\bf D62} 066004 (2000), {\it hep-th/9908056}.
\bibitem{Sakurai} J. J. Sakurai and S. F. Taun, {\sl Modern Quantum Mechanics} (Addison-Wesley Pub Co, 1994).
\bibitem{Chepelev} I. Chepelev and R. Roiban, {\sl JHEP} {\bf 0103} 001 (2001), {\it hep-th/0008090}.
\bibitem{Kos} I. Y. Aref'eva, D. M. Belov and A. S. Koshelev,
``A Note on UV / IR for Noncommutative Complex Scalar Field,'' {\it hep-th/0001215}.
\bibitem{Miao} L. Miao, ``Quantum corrections to noncommutative solitons,'' {\it hep-th/0011170}.
\bibitem{RS2} M. V. Raamsdonk and N. Seiberg, {\sl JHEP} {\bf 03} 035 (2000), {\it hep-th/0002186}.
\end{thebibliography}
\end{document}